\input{aipcheck}
\documentclass[
    ,final            % use final for the camera ready runs
  ]
  {aipproc}
\layoutstyle{6x9}
\begin{document}
\title{Phantom Crossing DGP Gravity}
%                \texttt{http://www.aip..org/pacs/index.html}>}
\author{Koichi Hirano}{
  address={Department of Physics, Ichinoseki National College of Technology, Ichinoseki 021-8511, Japan},
email={hirano@ichinoseki.ac.jp}
}
\author{Zen Komiya}{
  address={Department of Physics, Tokyo University of Science, Tokyo 162-8601, Japan}
}
%\author{<author3>}{
%  address={<common address for author2 and author3>}
%  ,altaddress={<author1 address>} % additional visiting address
%}
\begin{abstract}
We propose a phantom crossing Dvali--Gabadadze--Porrati (DGP) model. In our model, the effective equation of state of the DGP gravity crosses the phantom divide line. We demonstrate crossing of the phantom divide does not occur within the framework of the original DGP model or the DGP model developed by Dvali and Turner. By extending their model, we construct a model that realizes crossing of the phantom divide. 
%We find that the smaller the value of the new introduced parameter $\beta$ is, the older epoch crossing of the phantom divide occurs in. 
DGP models can account for late-time acceleration of the universe without dark energy. Phantom Crossing DGP model is more compatible with recent observational data from Type Ia Supernovae (SNIa), Cosmic Microwave Background (CMB) anisotropies, and Baryon Acoustic Oscillations (BAO) than the original DGP model or the DGP model developed by Dvali and Turner.
\end{abstract}

\maketitle

%%%%%%%%%%%%%%%%%%%%%%%%%%%%%%%%%%%%%%%%%%%%
%% MAINMATTER
%%%%%%%%%%%%%%%%%%%%%%%%%%%%%%%%%%%%%%%%%%%%

\section{INTRODUCTION}
Late time accelerated expansion of the universe was indicated by measurements of Type Ia supernovae (SNIa). It is not possible to account for this phenomenon within the framework of general relativity containing only matter and radiation. Therefore, a number of models containing "dark energy" have been proposed as the mechanism for the acceleration. However, the cosmological constant, which is the standard candidate for dark energy, cannot be explained by current particle physics due to its very small value.

The Dvali--Gabadadze--Porrati (DGP) model \cite{dva2000} is an extra dimension scenario. In this model, the universe is considered to be a brane; i.e., a four-dimensional (4D) hypersurface, embedded in a five-dimensional (5D) Minkowski bulk. On large scales, the late-time acceleration is driven by leakage of gravity from the 4D brane into 5D spacetime. Naturally, there is no need to introduce dark energy.
The action is
\begin{equation}
S=\frac{1}{16\pi}M_{(5)}^3\int_{bulk}{d^5x\sqrt{-g_{(5)}}R_{(5)}}+\frac{1}{16\pi}M_{(4)}^2\int_{brane}{d^4x\sqrt{-g_{(4)}}(R_{(4)}+L_m)}.
\end{equation}
A Friedmann-like equation on the brane is obtained as
\begin{equation}
H^2 = \frac{8\pi G}{3}\rho+\frac{H}{r_c} \label{dgp_fri}.
\end{equation}

Dvali and Turner \cite{dva2003} phenomenologically extended the Friedmann-like equation (Eq. \eqref{dgp_fri}) of the DGP model. This model interpolates between the original DGP model and the pure $\Lambda$CDM model with an additional parameter $\alpha$. The modified Friedmann-like equation is
\begin{equation}
H^2 = \frac{8\pi G}{3}\rho+\frac{H^{\alpha}}{{r_c}^{2-\alpha}}. \label{dt_fri}
\end{equation}
For $\alpha = 1$, this agrees with the original DGP Friedmann-like equation, while $\alpha = 0$ leads to an expansion history identical to that of $\Lambda$CDM cosmology.

According to various recent observational data including that of SNIa, it is possible that the effective equation of state parameter $w_{\rm eff}$, which is the ratio of the effective pressure $p_{\rm eff}$ to the effective energy density $\rho_{\rm eff}$, evolves from being larger than $-1$ to being less than $-1$; namely, it has currently crossed $-1$ (the phantom crossing).

\begin{figure}[h!]
  \includegraphics[height=.34\textheight,angle=270]{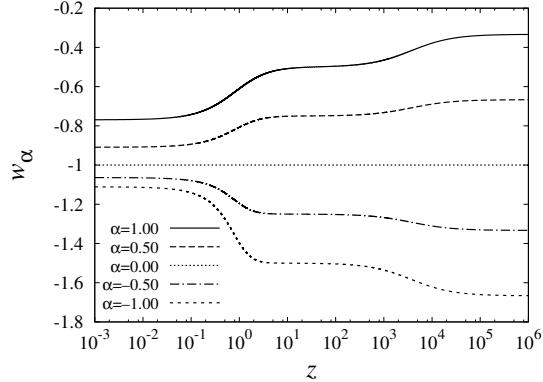}
  \caption{Effective equation of state $w_{\alpha}$, vs. redshift $z$ of the DGP gravity extended by Dvali and Turner for $\alpha = 1.00, 0.50, 0.00, -0.50$, and $-1.00$ (top to bottom) assuming $\Omega_{m,0} = 0.30$.}
  \label{fig:dt_wa}
\end{figure}

Fig. \ref{fig:dt_wa} shows a plot of the behavior of the effective equation of state of the DGP gravity represented by the second term on the right-hand side of Eq. (\ref{dt_fri}) of the DGP model by Dvali and Turner $w_\alpha$ versus the redshift $z$. When $\alpha$ is positive, the effective equation of state $w_\alpha$ will exceed $-1$ at all times. For negative $\alpha$, $w_\alpha$ is always less than $-1$. In the case of $\alpha = 0$, $w_\alpha$ is $-1$ constantly. Based on this analysis, crossing of the phantom divide does not occur in the DGP model extended by Dvali and Turner.
\section{Phantom crossing DGP model}
We propose the "Phantom Crossing DGP model" \cite{hir2009} that extends additionally the modified Friedmann equation (Eq. \eqref{dt_fri}) proposed by Dvali and Turner. Our model can realize crossing of the phantom divide line for the effective equation of state of the DGP gravity.

On the basis of the previous section's results, by change of the sign of $\alpha$ (from being positive to negative), the effective equation of state parameter varies from being larger than $-1$ to being less than $-1$.
Here, we shall assume that the physics that modifies Friedmann-like equation satisfies the following simple requirement:
\begin{itemize}
\item The sign of $\alpha$ of Eq. \eqref{dt_fri} varies from being positive to being negative {\bf keeping the model as simple as possible}. 
\end{itemize}
In accordance with this requirement, we make the following assumption, $\alpha = \beta - a$.
%\begin{equation}
%\alpha = \beta - a,
%\end{equation}
where $a$ is the scale factor (normalized such that the present day value is unity). The quantity $\beta$ is a constant parameter. In the period when the scale factor $a$ is less than the parameter $\beta$ ($\alpha > 0$), the effective equation of state exceeds $-1$. At the point when the scale factor $a$ equals $\beta$, ($\alpha = 0$), the equation of state's value will be $-1$. In the period when the scale factor $a$ exceeds the parameter $\beta$ ($\alpha < 0$), the equation of state will be less than $-1$. In this way, crossing of the phantom divide is realized in our model.

\begin{figure}[h!]
  \includegraphics[height=.31\textheight,angle=270]{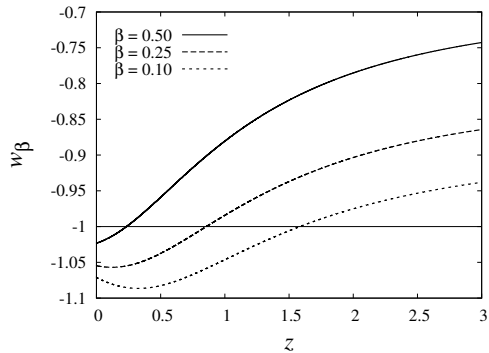}
  \caption{Effective equation of state of the DGP gravity of our model $w_{\beta}$, vs. redshift $z$. The solid, dashed, dotted lines represent the cases of $\beta = 0.50, 0.25$, and $0.10$, respectively (assuming $\Omega_{m,0} = 0.30$).}
  \label{fig:dgp_hirano}
\end{figure}

Replacing $\alpha$ by $\beta - a$ in Eq.(\ref{dt_fri}), the Friedmann-like equation in our model is given by
\begin{equation}
H^2 = \frac{8\pi G}{3}\rho+\frac{H^{\beta-a}}{{r_c}^{2-(\beta-a)}}. \label{hirano_fri}
\end{equation}
Fig. \ref{fig:dgp_hirano} shows a plot of the effective equation of state of the DGP gravity represented by the second term on the right-hand side of Eq. (\ref{hirano_fri}) of our model $w_\beta$ versus the redshift $z$. Our model realizes crossing of the phantom divide. 
%We find that the smaller the parameter $\beta$ is, the older epoch crossing of the phantom divide occurs in.
%%%%%%%%%%%%%%%%%%%%%%%%%%%%%%%%%%%%%%%%%%%%
%% Sample figure:
%%
%% The option [height=...] scales the picture to the given height,
%% without it it would be printed at its nominal size
%%%%%%%%%%%%%%%%%%%%%%%%%%%%%%%%%%%%%%%%%%%%

%\begin{figure}
%  \includegraphics[height=.3\textheight]{fig3}
%  \caption{Picture to fixed height}
%\end{figure}
\section{OBSERVATIONAL CONSTRAINTS}
We investigate the validity of our model in detailed comparison with recent observational data from SNIa, CMB anisotropies, and BAO.

\begin{table}[h!]
\begin{tabular}{lllrr}
\hline
Model & Best fit parameters & $\chi^2$ & $\Delta$AIC & $\Delta$BIC \\
\hline
Phantom Crossing DGP \cite{hir2009}  &  $\Omega_{m,0}=0.27$,~ $\beta=0.54$  &  246.729  &  ~~0.000  &  0.000 \\
DGP by Dvali and Turner \cite{dva2003}~~  &  $\Omega_{m,0}=0.27$,~ $\alpha=0.12$~~  &  247.086  &  ~~~0.357 & ~0.357 \\ 
Original DGP \cite{dva2000}  &  $\Omega_{m,0}=0.29$  &  269.002  &  20.273 & 16.603 \\
\hline
\caption{Comparison between models and recent observational data.\label{tab:kai2}}
\end{tabular}
\end{table}
 From the $\chi^2$ values presented in Table \ref{tab:kai2}, we can see that Phantom Crossing DGP model is more compatible with recent observations than the original DGP model or the DGP model developed by Dvali and Turner.
%%%%%%%%%%%%%%%%%%%%%%%%%%%%%%%%%%%%%%%%%%%%%%%%
%% BACKMATTER
%%%%%%%%%%%%%%%%%%%%%%%%%%%%%%%%%%%%%%%%%%%%%%%%

%\begin{theacknowledgments}
%  Infandum, regina, iubes renovare dolorem, Troianas ut opes et
%  lamentabile regnum cruerint Danai; quaeque ipse miserrima vidi, et
%  quorum pars magna fui. Quis talia fando Myrmidonum Dolopumve aut duri
%  miles Ulixi temperet a lacrimis?
%\end{theacknowledgments}
%%%%%%%%%%%%%%%%%%%%%%%%%%%%%%%%%%%%%%%%%%%%%%%%
%% The bibliography can be prepared using the BibTeX program or
%% manually.
%%
%% The code below assumes that BibTeX is used.  If the bibliography is
%% produced without BibTeX comment out the following lines and see the
%% aipguide.pdf for further information.
%%
%% For your convenience a manually coded example is appended
%% after the \end{document}
%%%%%%%%%%%%%%%%%%%%%%%%%%%%%%%%%%%%%%%%%%%%%%%%
%%%%%%%%%%%%%%%%%%%%%%%%%%%%%%%%%%%%%%%%%%%%%%%%
%% You may have to change the BibTeX style below, depending on your
%% setup or preferences.
%%
%%
%% For The AIP proceedings layouts use either
%%%%%%%%%%%%%%%%%%%%%%%%%%%%%%%%%%%%%%%%%%%%
\bibliographystyle{aipproc}   % if natbib is available
%\bibliographystyle{aipprocl} % if natbib is missing
%%%%%%%%%%%%%%%%%%%%%%%%%%%%%%%%%%%%%%%%%%%
%% You probably want to use your own bibtex database here
%%%%%%%%%%%%%%%%%%%%%%%%%%%%%%%%%%%%%%%%%%%
%\bibliography{sample}
%%%%%%%%%%%%%%%%%%%%%%%%%%%%%%%%%%%%%%%%%%%
%% Just a reminder that you may have to run bibtex
%% All of it up to \end{document} can be removed
%% if you don't like the warning.
%%%%%%%%%%%%%%%%%%%%%%%%%%%%%%%%%%%%%%%%%%%
%\IfFileExists{\jobname.bbl}{}
% {\typeout{}
%  \typeout{******************************************}
%  \typeout{** Please run "bibtex \jobname" to optain}
%  \typeout{** the bibliography and then re-run LaTeX}
%  \typeout{** twice to fix the references!}
%  \typeout{******************************************}
%  \typeout{}
% }
%%%%%%%%%%%%%%%%%%%%%%%%%%%%%%%%%%%%%%%%%%%
%% The following lines show an example how to produce a bibliography
%% without the help of the BibTeX program. This could be used instead
%% of the above.
%%%%%%%%%%%%%%%%%%%%%%%%%%%%%%%%%%%%%%%%%%%

%\endinput
\end{document}